\documentclass[aps,prl,preprint,superscriptaddress]{revtex4-1}
\usepackage{graphicx}

\newcommand {\unit} [1] {\; \mathrm {#1}}

\newcommand{\met}{\mbox{${\rm \not\! E}_{\rm T}$}}

\begin{document}

\title{Stealth Supersymmetry}

\author{JiJi Fan}
\affiliation{Department of Physics, Princeton University, Princeton, NJ, 08540}

\author{Matthew Reece}
\affiliation{Princeton Center for Theoretical Science, Princeton University, Princeton, NJ, 08540}

\author{Joshua T. Ruderman}
\affiliation{Department of Physics, Princeton University, Princeton, NJ, 08540}

\date{\today}

\begin{abstract}
We present a broad class of supersymmetric models that preserve $R$-parity but lack missing energy signatures. These models have new light particles with weak-scale supersymmetric masses that feel SUSY breaking only through couplings to the MSSM. This small SUSY breaking leads to nearly degenerate fermion/boson pairs, with small mass splittings and hence small phase space for decays carrying away invisible energy.  The simplest scenario has low-scale SUSY breaking, with missing energy only from soft gravitinos. This scenario is natural, lacks artificial tunings to produce a squeezed spectrum, and is consistent with gauge coupling unification. The resulting collider signals will be jet-rich events containing false resonances that could resemble signatures of $R$-parity violation.  We discuss several concrete examples of the general idea, and emphasize $\gamma j j$ resonances, displaced vertices, and very large numbers of $b$-jets as three possible discovery modes.
\end{abstract}


\maketitle

\section{Introduction} 
The Large Hadron Collider (LHC) has embarked on a broad campaign to discover weak scale supersymmetry (SUSY, see~\cite{Martin:1997ns} for a review).  Many SUSY searches are now underway, in hopes of discovering energetic jets, leptons, and/or photons produced by the decays of superpartners.  A common feature of most SUSY searches~\cite{LHCjetsmet} is that they demand a large amount of missing transverse energy $(\met)$ as a strategy to reduce Standard Model (SM) backgrounds.  This approach is motivated by $R$-parity, which, if preserved, implies that the lightest superpartner (LSP) is stable and contributes to missing energy.  In this paper, we introduce a new class of SUSY models that preserve $R$-parity, yet lack missing energy signatures.  These models of {\it Stealth Supersymmetry} will be missed by standard SUSY searches. 

Even when $R$-parity is preserved, the lightest SM (`visible' sector) superpartner (LVSP) can decay, as long as there is a lighter state that is charged under $R$-parity.  This occurs, for example, when SUSY is broken at a low scale (as in gauge mediated breaking, reviewed by~\cite{Giudice:1998bp}), and the LVSP can decay to a gravitino, which is stable and contributes to missing energy.  Here, we consider the additional possibility that there exists a new hidden sector of particles at the weak scale, but lighter than the LVSP\@.  If SUSY is broken at a low scale, it is natural for the hidden sector to have a spectrum that is approximately supersymmetric, with a small amount of SUSY breaking first introduced by interactions with SM fields.

\begin{figure}[!h]
\includegraphics[width=0.9\columnwidth]{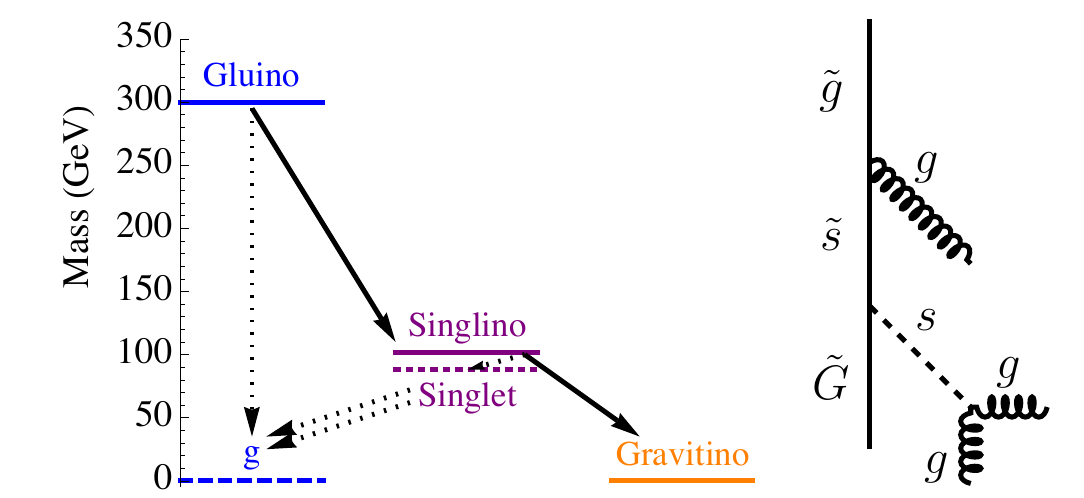}
\caption{An example spectrum and decay chain for Stealth SUSY with gluino LVSP.\label{fig:schema}}\end{figure}

The generic situation described above is all that is required to suppress missing energy in SUSY cascades.  The LVSP can decay into a hidden sector field, $\tilde X$,
which we take to be fermionic, and heavier than its scalar superpartner, $X$.  Then, $\tilde X$ decays to a stable gravitino and its superpartner, $\tilde X \rightarrow \tilde G X$, and $X$, which is even under $R$-parity, can decay back to SM states like jets, $X \rightarrow j j$.  Because the spectrum in the hidden sector is approximately supersymmetric, the mass splitting is small within the $X$ supermultiplet, $m_{\tilde X} - m_X \ll m_{\tilde X}$.  Therefore, there is no phase space for the gravitino to carry momentum: the resulting gravitino is soft and missing energy is greatly reduced.  We illustrate the spectrum, and decay path, in figure~\ref{fig:schema}, where $X$ and $\tilde X$ correspond to a singlet and singlino.  We emphasize that this scenario requires no special tuning of masses. The approximate degeneracy between $X$  and $\tilde X$ is enforced by a symmetry: supersymmetry!

Two assumptions can be relaxed. First, a fermion other than the gravitino can end the cascade, if its mass fits in the small available phase space: we can generalize to $\tilde{X} \to \tilde{N}X$ for a variety of light neutral fermions $\tilde{N}$. Because gravitino couplings are $1/F$-suppressed, such decays are often preferred if available. Then, we need not assume low-scale SUSY breaking; gravity mediation can also give rise to this scenario, if a suppressed SUSY-breaking splitting between $\tilde{X}$ and $X$ is natural. This calls for sequestering, an idea that already plays a key role in such scenarios as anomaly mediation~\cite{AMSB}.

A hidden sector may therefore eliminate missing energy, making the  SUSY searches ineffective at the LHC.  Moreover, the LEP and Tevatron limits on supersymmetry mostly rely on missing energy, and do not apply to these models.  This raises the interesting possibility of {\it hidden} SUSY: superpartners may be light enough to have been produced copiously at LEP and the Tevatron, yet missed, because their decays do not produce missing energy.  Our proposal is morally similar, but more far reaching, than the idea that the higgs boson may be light, but hidden from LEP by exotic decay modes (see the references within~\cite{Chang:2008cw}, and more recently~\cite{HidingHiggs}). It also has a great deal in common with SUSY models containing Hidden Valleys~\cite{Strassler:2006qa}, though in previous discussions $\met$ has been suppressed by longer decay chains, rather than supersymmetric degenerate states. Fortunately, there are a number of experimental handles that can be used to discover stealth supersymmetry.  Possible discovery modes that we emphasize in this paper include highly displaced vertices, triple resonances such as $\gamma j j$, and the presence of a very large number of $b$-jets.

\section{Models} There are many possible implementations of stealth SUSY.  To illustrate the simplicity of our idea, we present two minimal models, where the hidden sector consists of one singlet chiral superfield, $S$, with a supersymmetric mass, $W \supset m/2 \, \, S^2$.  In the first example, $S$ couples to SM higgses, $S H_u H_d$, and sees SUSY breaking at tree-level.  In the second example, $S$ couples to a messenger field, $Y$, charged under the SM gauge symmetries, $S Y \bar Y$, and experiences SUSY breaking at one-loop.
\vspace{-0.7cm}

\subsection{\bf $\mathbf{S H_u H_d}$}
We add a singlet chiral superfield $S$, as in~\cite{Drees:1988fc}, but suppose it feels SUSY breaking only through interactions with the MSSM higgs:
\begin{equation}
W = \frac{m}{2} S^2 + \frac{\kappa}{3} S^3 + \lambda S H_u H_d + \mu H_u H_d.
\end{equation}
In the limit of small $\lambda$, the model has vacua near $\left<S\right> = 0$ and $\left<S\right> = -m/\kappa$, either of which can be lower depending on the parameters. Given small $\lambda$ and $\kappa$, the mass splitting in $S$ is $\sim \lambda \kappa \mu v^2/m^2$. The lightest field in the $S$ multiplet can be the scalar $s$ or pseudoscalar $a$, depending on regions of parameter space considered. The (pseudo)scalar decays dominantly to $b\bar{b}$ through mixing with the higgs. A benchmark point is shown in Table~\ref{tab:shuhd}.

\begin{table}
\begin{tabular}{|c|c|}
\hline
\multicolumn{2}{|c|}{$S H_u H_d$} \\
\hline
$m = 80\unit{GeV}$ & $m_a = 90\unit{GeV} \quad m_s = 103\unit{GeV}$  \\
$ \mu = 300\unit{GeV}$ &  $m_h = 125 \unit{GeV}$ \\
$\lambda = -0.02 \quad \kappa = 0.5$ & $\sigma_{s Z} = 0.22 \, \sigma_{h Z}$\\
$\tan \beta = 10 \quad m_A = 700\unit{GeV}$ &$ \Gamma_a = 6 \times 10^{-8}\unit{GeV}$ \\
$M_1 = 200\unit{GeV}$ &  $m_{\tilde s} = 100 \unit{GeV}$\\
$M_2 = 300\unit{GeV}$ & $N_{\tilde s (\tilde H_u, \tilde H_d)} = (-0.014, 0.0059)$ \\
$M = -2\unit{TeV}$& $N_{\tilde s( \tilde B, \tilde W^0)} = (0.0063, - 0.0058)$\\
\hline
\end{tabular}
\caption{A benchmark point for the $S H_u H_d$ model. To lift the Higgs mass above the experimental limit (even if stops are light), we add $\left(H_u H_d\right)^2/M$ to the superpotential~\cite{Dine:2007xi}. Note that $e^+e^- \to Zs( \to \bar{b}b)$ could be consistent with the 2$\sigma$ excess at $\approx 100$ GeV observed by LEP~\cite{Schael:2006cr}.}
\label{tab:shuhd}
\end{table}

\subsection{\bf $\mathbf{S Y\bar{Y}}$}
This scenario involves two more chiral supermultiplets $Y$ and $\bar{Y}$ in the ${\bf 5}$ and ${\bf {\bar 5}}$ of $SU(5)_{GUT}$. We consider a superpotential:
\begin{equation}
W = \frac{m}{2} S^2 + \lambda S Y \bar{Y} + m_Y Y \bar{Y}.
\end{equation}
Here $m_Y$ and $m$ are supersymmetric masses, with $m_Y \sim $ TeV and $m \sim 100$ GeV. Soft masses $\tilde{m}_D^2,\tilde{m}_L^2$ for the ${\bf 3}$ and ${\bf 2}$ in $Y$ (and equal ones for ${\bar Y}$) are generated by gauge mediation and through RG running lead to a negative soft mass-squared for the scalar $s$
\begin{equation}
m_s^2 \sim -\frac{|\lambda|^2}{(4\pi)^2} \left(6 \tilde{m}_D^2 + 4 \tilde{m}_L^2\right) \log\frac{M_{\rm mess}^2}{m_Y^2}.
\end{equation}
For $\tilde{m}_D, \tilde{m}_{L} \sim {\cal{O}}(1$ TeV), this leads to splittings of order 10 GeV or less with reasonable choices of couplings and scales.

Integrating out $Y$ and $\bar{Y}$ at one loop yields operators such as $\lambda^a  \sigma_{\mu\nu}G^{a\mu\nu}\tilde{s}$ and $sG^a_{\mu\nu}G^{a\mu\nu}$. These interactions would induce decays of the gluino to singlino plus gluon and of the scalar $s$ to gluons, as in Fig.~\ref{fig:schema}. Similar operators between $S$ and other SM  vector multiplets exist, which allow decays of neutralinos (charginos) to singlino plus $\gamma/Z$ ($W$) and of $s$ to two $\gamma$'s. A benchmark point is shown in Table~\ref{tab:syy}.

Finally we comment that the supersymmetric mass of $S$ could arise dynamically through retrofitting, which can also be related to the SUSY-breaking scale~\cite{Dine:2006gm}. Global symmetries can be arranged to forbid large SUSY breaking for $S$ that would spoil our picture.

\begin{table}
 \begin{tabular}{| c | c |}
\hline
\multicolumn{2}{|c|}{$S Y \bar Y$} \\
\hline
$m = 100\unit{GeV}$ & $m_{\tilde s} = 100\unit{GeV}$ \\
$\lambda = 0.2$ &$m_{s,a} = 91\unit{GeV}$ \\
$m_{Y} = 1000 \unit{GeV} $&  $\Gamma_{s,a} = 2 \times 10^{-7}\unit{GeV} $ \\
$\tilde{m}_{D} = 300 \unit{GeV} \quad \tilde{m}_{L} = 200\unit{GeV}$&$\mathrm{Br}_{s,a \rightarrow \gamma \gamma}= 4 \times 10^{-3}$\\
$M_{\rm mess} = 100\unit{TeV}$ & \\
\hline
 \end{tabular}
 \caption{A benchmark point for the $S Y {\bar Y}$ model.}
 \label{tab:syy}
\end{table}

\subsection{Further Possibilities}
Similar spectra and phenomenology could be achieved in many other models. One next-to-minimal possibility is to add to the MSSM a vector superfield $V$, which is associated with a $U(1)^\prime$ spontaneously broken at the weak scale. SM fields transforming under the $U(1)^\prime$ would transmit the SUSY breaking to $V$. Like the $S Y\bar{Y}$ model, the mass splitting is of order ${\cal{O}}(m_{soft}/(4\pi)^2)\sim {\cal{O}}$ (10 GeV). A similar generalization could involve supersymmetric vectorlike confinement~\cite{Kilic:2009mi}.  Even the MSSM may include a form of stealth SUSY, if there is an approximate degeneracy between the right-handed stop and top masses.  The stop can decay to a top plus a soft gravitino or light bino, which may obscure light SUSY in top backgrounds.  This scenario is natural when $m_{\bar {u}_3}^2 \ll m_t^2$ and the stop mixing is small. 

As mentioned in the Introduction, we can relax the gravitino LSP assumption. An SM singlet chiral superfield $N$ kept light by a symmetry, with $S^2 N$ in the superpotential, allows $\tilde{s} \to s \tilde{N}$. Such generalizations allow for prompt decays even when the SUSY breaking scale is not small. Extending to gravity mediation requires sequestering of $F/M_{Pl}$ corrections to $m_s^2$, which is a model-building complication we leave for future work. 

\section{2 Body vs 3 Body} 
\begin{figure}[!h]
\includegraphics[width=0.9\columnwidth]{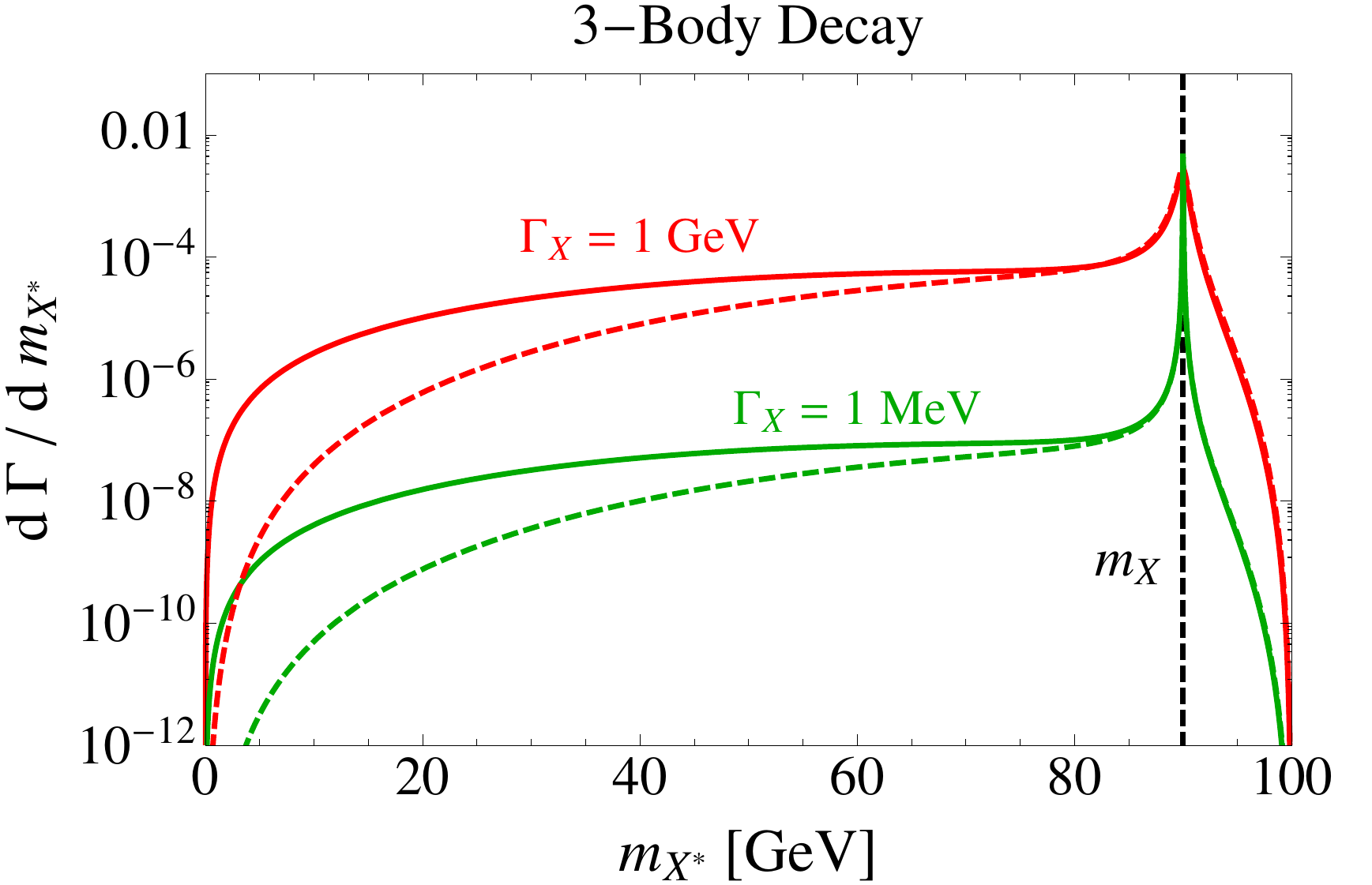}
\caption{Three-body decays $\tilde{X} \to \tilde{G}X^*(\to j j)$ can become important if $\delta m$ is small and $\Gamma_X$ is not too small. The solid line is for $X$ decaying through a Yukawa coupling, whereas the dashed line is for $X$ decaying to gauge bosons.\label{fig:threebody}}
\end{figure}

The decay width of $\tilde{X} \to X \tilde{G}$ (at $\delta m \ll m$ and neglecting mixings) is given by~\cite{GMSBwidths}:
\begin{equation}
\Gamma_{\tilde{X}}=\frac{m_{\tilde X}^5}{16\pi F^2} \left(1 - \frac{m_X^2}{m_{\tilde X}^2}\right)^4 \approx \frac{m_{\tilde{X}} \, (\delta m)^4}{\pi F^2} \, .
\label{eq:smallwidth}
\end{equation}
For SUSY breaking scale $\sqrt F= 100\, {\rm{TeV}}$, $m_{\tilde{X}}=$100 GeV and $m_X =$ 90 GeV, the decay length is 8 cm. In addition to the $F^2$ suppression of any decay to gravitino, the small mass splitting further suppresses the two-body decay while it enhances the branching ratio of the three-body decay through an off-shell $X$, $\tilde{X} \to \tilde{G}X^*(\to j j)$, which has differential width~\cite{GMSBwidths}:
\begin{equation}
\frac{d\Gamma}{dq^2} \propto \frac{q^{2n} \left(q^2 - m_{\tilde X}^2\right)^4}{\left(q^2 - m_X^2\right)^2 + m_X^2 \Gamma_X^2},
\label{eq:3bodywidth}
\end{equation}
where $n = 1$ for a decay through a Yukawa coupling $X\bar{\psi}\psi$, whereas $n=2$ for $X$ decaying through an operator $XG^a_{\mu\nu}G^{a\mu\nu}$. In this case, the missing energy could be increased. However, as we show in Figure~\ref{fig:threebody}, as long as $\Gamma_X$ is small, the two-body decay will always dominate. 

\section{Have You $\met$, SUSY?} 
\begin{figure}[!h]
\includegraphics[width=0.9\columnwidth]{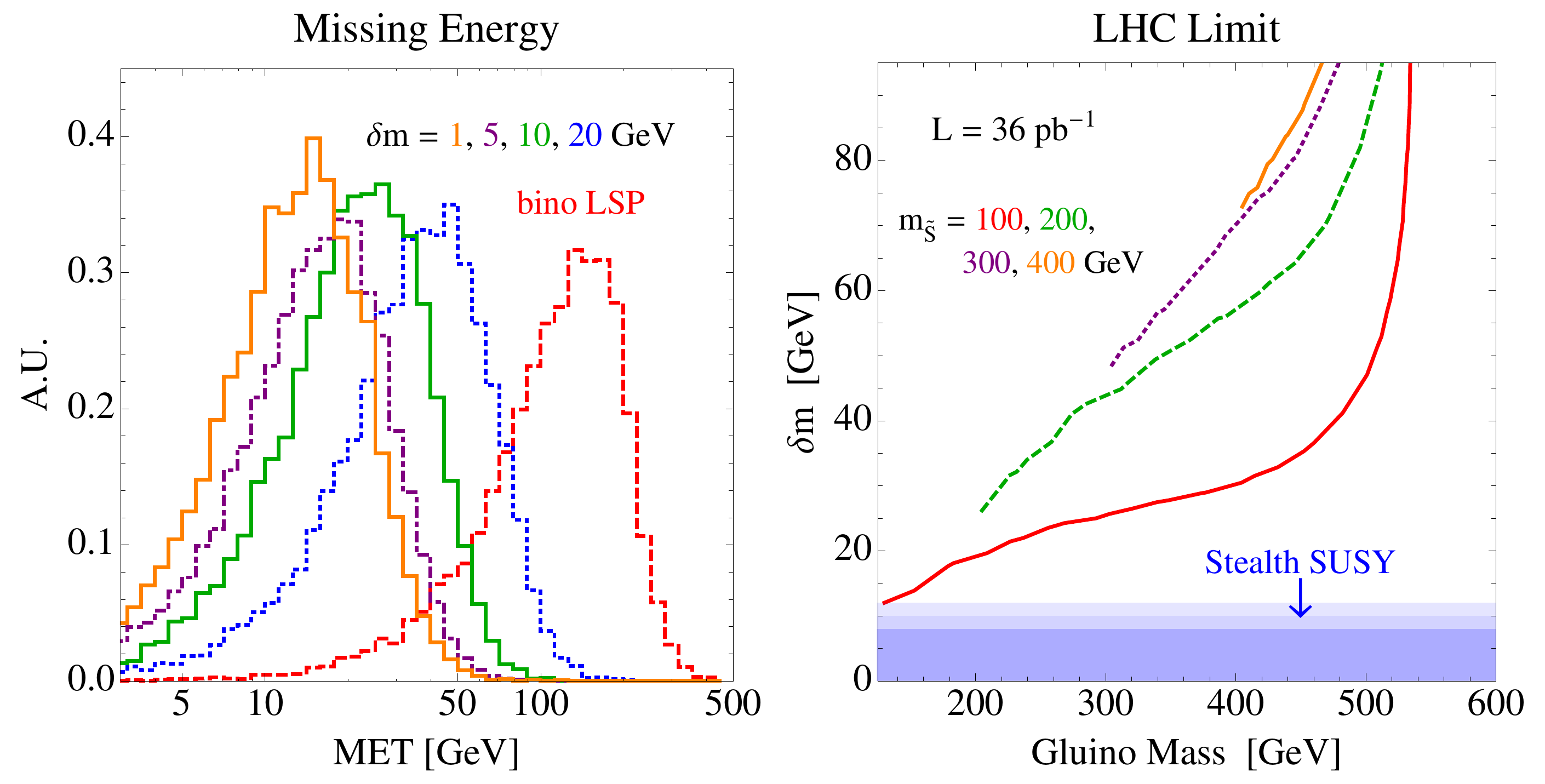}
\caption{At left: Missing transverse energy (MET) in a SUSY scenario with gluino decaying to bino ($\tilde{g} \to q\bar{q}\tilde{\chi}^0_1$) compared to a decay chain $\tilde{g} \to g(\tilde{s}\to \tilde{G}(s\to gg))$ as in Fig.~\ref{fig:schema}. The curves are labeled by mass splitting. At right: estimated exclusion contours using refs.~\cite{LHCjetsmet}. The region above the curves is excluded. The curves are labeled by singlino mass and begin where $m_{\tilde g} > m_{\tilde s}$.\label{fig:metdemo}}
\end{figure}

Missing energy is dramatically reduced by small mass splittings. The gravitino LSP, in the rest frame of a decaying singlino, has momentum $\delta m \ll m_{\tilde s}$, implying a smaller $\met$ in the lab frame. Figure~\ref{fig:metdemo} shows the $\met$ distributions for a 300 GeV gluino decaying in a standard SUSY scenario to a stable 100 GeV bino, versus decaying through a 100 GeV singlino with the singlet at various masses. This simulation used Pythia 6.4~\cite{Sjostrand:2006za}, BRIDGE~\cite{Meade:2007js}, and PGS as a simple detector model~\cite{PGS}. A splitting of 10 GeV reduces the $\met$ by an order of magnitude, while a 1 GeV splitting saturates the $\met$ reduction, as jet mismeasurement becomes the dominant effect. For longer lifetimes, an additional source of $\met$ arises when the momentum and the vector from the origin to the calorimeter are no longer aligned. We modeled this effect and found that it adds to the tail of $\met$ distributions, but is a very small effect for 10 cm lifetimes and only moderately important at 50 cm lifetimes. We have modeled several ATLAS and CMS searches~\cite{LHCjetsmet}, and present the strongest estimated limits on $\tilde{g}\tilde{g}$ production with $\tilde{g}\to g\tilde{s}$, $s\to gg$ in Fig.~\ref{fig:metdemo}.

\section{Detecting Stealth SUSY} 
While standard missing $E_T$ searches overlook stealth SUSY, a variety of experimental handles exist. The small width (Eq.~\ref{eq:smallwidth}) gives rise to displacements of millimeters, centimeters, or more in stealth SUSY events. High multiplicity final states can also be an interesting general search strategy~\cite{HighMultiplicity}. Other signatures are more model-dependent. In the $SH_u H_d$ model, $s$ dominantly decays to $b\bar{b}$, so that most events will include at least four $b$'s. Because the singlino mixes with higgsinos, other decays like $\tilde{B} \to \tilde{s}h(\to b\bar{b})$ can occur to produce more $b$'s. In fact, a chain $\tilde{g} \to \tilde{b} \to \tilde{B} \to \tilde{s}$ can produce as many as 12 $b$'s in a single event! The displaced vertex and $b$-jet signatures of stealth SUSY resemble aspects of Hidden Valley phenomenology~\cite{HiddenValleyRefs}. In the $SY\bar{Y}$ model, the colored $Y$ fields may decay only through GUT-suppressed operators, opening the possibility of long-lived, $R$-hadron-like phenomenology~\cite{Luty:2010vd}. The uncolored fields in $Y$ could be a candidate for dark matter, if direct detection through a $Z$ is forbidden by an inelastic splitting~\cite{Hall:1997ah}.

\subsection{False Resonances} 
Because the gravitino is soft, invariant masses made of visible particles may reconstruct peaks for $R$-odd particles. For instance, the decay chain
\begin{equation}
\tilde{q} \to q (\tilde{B} \to \gamma (\tilde{s} \to \tilde{G} (s \to gg)))
\label{eq:squarkchain}
\end{equation}
will have $M(\gamma g g) \approx m_{\tilde{B}}$ and $M(\gamma g g q) \approx m_{\tilde{q}}$. We advocate searching for resonances composed of a gauge boson ($\gamma,Z,W^\pm$) and a pair of jets, to reconstruct bino and wino (co-)LVSPs. (A $\gamma j j$ resonance has been discussed in pure-glue Hidden Valleys~\cite{Juknevich:2009ji}.) A gluino decay chain, $\tilde{g} \to g\tilde{s} \to ggg\tilde{G}$, can lead to a 3-jet resonance, which strongly resembles the $\tilde{g} \to 3 q$ decay in R-parity violating SUSY with $UDD$ couplings. An approach to overcoming combinatorics to find $jjj$ resonances based on cuts in the $(M(jjj), \sum_j p_T)$ plane has been pursued by CDF and CMS~\cite{RPVjjj}. Our simulations show that these techniques have similar reach for our gluino LVSP case.

\begin{figure}
\includegraphics[width=0.7\columnwidth]{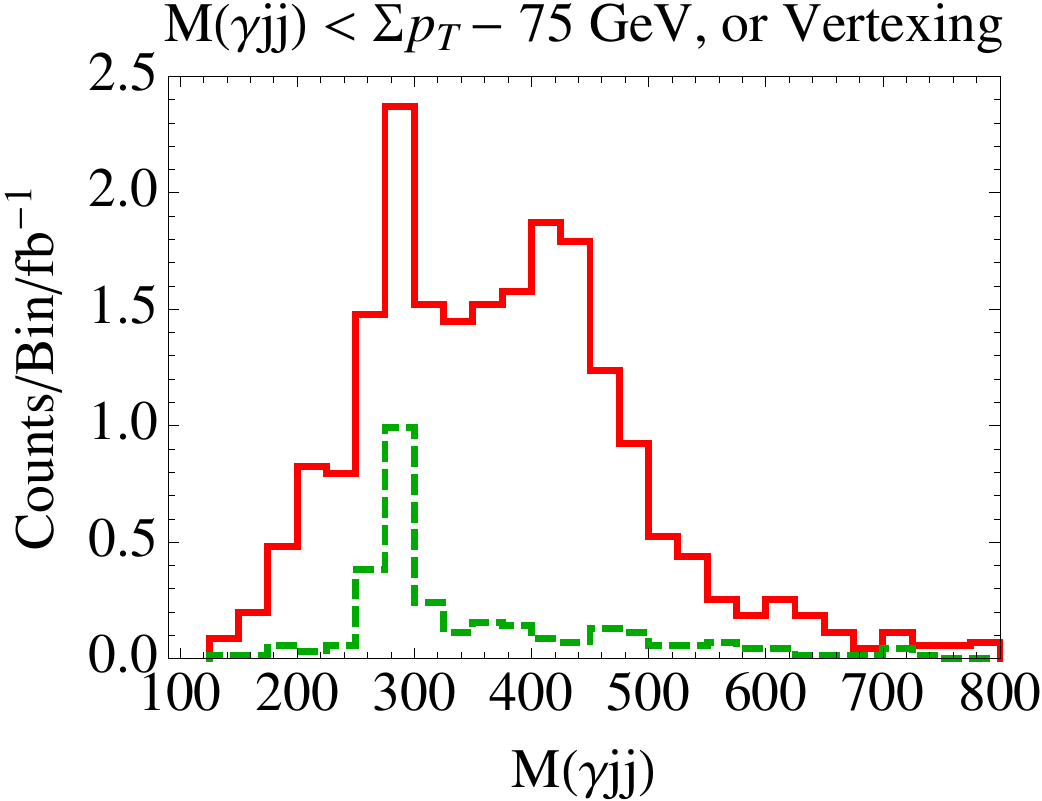}
\caption{Red, solid: $M(\gamma j j)$ for all triplets passing a cut on mass relative to $\sum p_T$. Note the peak at the bino mass of 300 GeV and the falling feature near the squark mass of 500 GeV. Green, dashed: $M(\gamma j j)$ for jet pairs tagged as an $s$ decay via vertexing, together with each photon (no cut on $\sum p_T$ is applied).\label{fig:photonjet}}
\end{figure}

We have studied squark pair production, with decays as in Eq.~\ref{eq:squarkchain}, in more detail. We fix a benchmark point: $m_{\tilde{q}} = 500$ GeV for $\tilde{u}_R$ and $\tilde{c}_R$ with other squarks decoupled, $m_{\tilde{B}} = 300$ GeV, $m_{\tilde{s}} = 100$ GeV, and $m_s = 90$ GeV. We generate events with Pythia and a modified decay table, and reconstruct jets using FastJet's anti-$k_t$ algorithm ($R=0.5$)~\cite{FastJet}. Studies of $\gamma\gamma+{\rm jets}$ backgrounds (using MadGraph 4~\cite{MadGraph} with MLM matching~\cite{MLM}, Pythia, and comparisons to a recent CMS study~\cite{Chatrchyan:2011jx} that measured $\gamma\gamma$ distributions) show that requiring two photons with $E_T > 120$ GeV and  $|\eta| < 1.44$, along with at least two jets with $E_T > 45$ GeV and $\sum_{jets} E_T > 200$ GeV, reduces the background below the signal. The challenge is then combinatorics. We apply the technique of Refs.~\cite{RPVjjj}: we find that forming all $\gamma jj$ triplets and requiring $M(\gamma jj) < \sum_{\gamma,j,j} p_T - 75$ GeV brings out features from the combinatoric background. The distribution is shown in Fig.~\ref{fig:photonjet}; its broad outline would become apparent with around 1 fb$^{-1}$ of data, while several fb$^{-1}$ would be needed to make the structure clear. The expected feature is present at the bino mass, $M(\gamma j j)\approx 300$ GeV, but there is also an apparent edge or endpoint near the squark mass, $M(\gamma jj)\approx 500$ GeV. In fact, we find that many jets have substructure from $s \to gg$ reconstructed as one jet, a ``singlet jet."  With sufficient data (tens of fb$^{-1}$) the bino is also visible in $M(\gamma j)$ for massive jets. This resembles using substructure to reconstruct neutralinos and squarks in R-parity violating SUSY~\cite{RPVsub}. In the $SH_uH_d$ model, the ``singlet jet" contains two $b$ subjets which could be searched for by the same techniques as proposed to discover a boosted light higgs~\cite{boostedhiggs}.
We also show in Fig.~\ref{fig:photonjet} that if displaced vertices are used to select jet pairs from the same decay, combinatoric effects can be greatly reduced. We leave more detailed studies of event reconstruction and jet substructure for future work. 

\section{Remarks} 
Stealth SUSY gives a natural escape for supersymmetry from $\met$ searches, but it can be detected through displaced vertices, $b$-jet multiplicities, and mass peaks reconstructed from gauge bosons and jets. Measurements of such events could reveal the existence of a nearly degenerate fermion and boson. (The idea that supersymmetry may first be discovered in a hidden sector appears in~\cite{Strassler:2008bv}.) One simple example is production of a sbottom LVSP in the $SH_u H_d$ model. The techniques of~\cite{RPVjjj} plus $b$-tagging could find 3$b$ resonances, but with a production cross section consistent with a colored scalar at the measured mass. (Tagging of $b$ jets is complicated by the fact that even non-$b$ jets will be displaced in these models, but soft lepton tags could still be useful on a subset of events.) The inconsistency between the apparent production of a scalar and a measured fermionic final state would be a strong signal that a soft fermion has escaped. We will take a closer look at techniques for confirming the spins of $\tilde{s}$ and $s$ and establishing this smoking gun for supersymmetry in future work.

{\bf Acknowledgments.} We thank Nima Arkani-Hamed, Cliff Cheung, John Paul Chou, Pavel Fileviez P\'erez, Eva Halkiadakis, David E. Kaplan, Amit Lath, Martin Schmaltz, Scott Thomas, and Neal Weiner for useful discussions. J.F. is supported by the DOE grant DE-FG02-91ER40671. M.R. thanks the PCTS for its support.


\begin{thebibliography}{nn}

\bibitem{Martin:1997ns}
  S.~P.~Martin,
  arXiv:hep-ph/9709356.
 
\bibitem{LHCjetsmet}    
  V.~Khachatryan {\it et al.}  [CMS Collaboration],
  Phys.\ Lett.\  B {\bf 698}, 196 (2011);
  [arXiv:1101.1628 [hep-ex]].
  J.~B.~G.~da Costa {\it et al.}  [Atlas Collaboration],
  arXiv:1102.5290 [hep-ex];
 CMS, 
 PAS-SUS-10-005, 2011;
 CMS, 
 PAS-SUS-10-009, 2011.
 
\bibitem{Giudice:1998bp}
  G.~F.~Giudice and R.~Rattazzi,
  Phys.\ Rept.\  {\bf 322}, 419 (1999).
  [arXiv:hep-ph/9801271].
 
\bibitem{AMSB}
  L.~Randall and R.~Sundrum,
  Nucl.\ Phys.\  B {\bf 557}, 79 (1999).
  [arXiv:hep-th/9810155].
  
\bibitem{Chang:2008cw}
  S.~Chang, R.~Dermisek, J.~F.~Gunion, N.~Weiner,
  Ann.\ Rev.\ Nucl.\ Part.\ Sci.\  {\bf 58}, 75-98 (2008).
  [arXiv:0801.4554 [hep-ph]].
  
\bibitem{HidingHiggs}  
  B.~Bellazzini, C.~Csaki, A.~Falkowski and A.~Weiler,
  Phys.\ Rev.\  D {\bf 80}, 075008 (2009);
  [arXiv:0906.3026 [hep-ph]];
  B.~Bellazzini, C.~Csaki, A.~Falkowski and A.~Weiler,
  Phys.\ Rev.\  D {\bf 81}, 075017 (2010);
  [arXiv:0910.3210 [hep-ph]];
  A.~Falkowski, J.~T.~Ruderman, T.~Volansky and J.~Zupan,
  JHEP {\bf 1005}, 077 (2010).
  [arXiv:1002.2952 [hep-ph]].
  
\bibitem{Strassler:2006qa}
  M.~J.~Strassler,
  arXiv:hep-ph/0607160.
  
\bibitem{Drees:1988fc}
  M.~Drees,
  Int.\ J.\ Mod.\ Phys.\  A {\bf 4}, 3635 (1989).
  
\bibitem{Dine:2007xi}
  M.~Dine, N.~Seiberg and S.~Thomas,
  Phys.\ Rev.\  D {\bf 76}, 095004 (2007).
  [arXiv:0707.0005 [hep-ph]].
  
\bibitem{Schael:2006cr}
  S.~Schael {\it et al.}  [ALEPH Collaboration and DELPHI Collaboration and
                  L3 Collaboration and OPAL Collaboration],
  Eur.\ Phys.\ J.\  C {\bf 47}, 547 (2006).
  [arXiv:hep-ex/0602042].
  
\bibitem{Dine:2006gm}
  M.~Dine, J.~L.~Feng and E.~Silverstein,
  Phys.\ Rev.\  D {\bf 74}, 095012 (2006);
  [arXiv:hep-th/0608159];
  D.~Green, T.~Weigand,
  [arXiv:0906.0595 [hep-th]].  
  
\bibitem{Kilic:2009mi}
  C.~Kilic, T.~Okui and R.~Sundrum,
  JHEP {\bf 1002}, 018 (2010).
  [arXiv:0906.0577 [hep-ph]].
  
\bibitem{GMSBwidths}
  S.~Ambrosanio, G.~L.~Kane, G.~D.~Kribs, S.~P.~Martin and S.~Mrenna,
  Phys.\ Rev.\  D {\bf 54}, 5395 (1996).
  [arXiv:hep-ph/9605398].
  
\bibitem{Sjostrand:2006za}
  T.~Sjostrand, S.~Mrenna and P.~Z.~Skands,
  ``PYTHIA 6.4 Physics and Manual,''
  JHEP {\bf 0605}, 026 (2006).
  [arXiv:hep-ph/0603175].
  
\bibitem{Meade:2007js}
  P.~Meade and M.~Reece,
  arXiv:hep-ph/0703031.
  
    \bibitem{PGS}
J. Conway, PGS: \url{http://physics.ucdavis.edu/~conway/research/software/pgs/pgs4-general.htm}

\bibitem{HighMultiplicity}
M.~Lisanti, P.~Schuster, M.~Strassler, and N.~Toro, to appear.

\bibitem{HiddenValleyRefs}
  M.~J.~Strassler and K.~M.~Zurek,
  Phys.\ Lett.\  B {\bf 651}, 374 (2007).
  [arXiv:hep-ph/0604261];
  M.~J.~Strassler and K.~M.~Zurek,
  Phys.\ Lett.\  B {\bf 661}, 263 (2008).
  [arXiv:hep-ph/0605193].

\bibitem{Luty:2010vd}
  M.~A.~Luty, D.~J.~Phalen and A.~Pierce,
  Phys.\ Rev.\  D {\bf 83}, 075015 (2011).
  [arXiv:1012.1347 [hep-ph]].
  
\bibitem{Hall:1997ah}
  L.~J.~Hall, T.~Moroi and H.~Murayama,
  Phys.\ Lett.\  B {\bf 424}, 305 (1998).
  [arXiv:hep-ph/9712515].
  
\bibitem{Juknevich:2009ji}
  J.~E.~Juknevich, D.~Melnikov and M.~J.~Strassler,
  JHEP {\bf 0907}, 055 (2009).
  [arXiv:0903.0883 [hep-ph]].

\bibitem{RPVjjj}  
  R.~Essig, Ph.D. Thesis, Rutgers University, 2008;
   T.~Aaltonen et al.  [CDF Collaboration],
  arXiv:1105.2815 [hep-ex];
  CMS Collaboration, 
 PAS EXO-11-001, 2011.

\bibitem{FastJet}
  M.~Cacciari and G.~P.~Salam,
  Phys.\ Lett.\  B {\bf 641}, 57 (2006);
  [arXiv:hep-ph/0512210];
  M.~Cacciari, G.~P.~Salam and G.~Soyez,
  JHEP {\bf 0804}, 063 (2008);
  [arXiv:0802.1189 [hep-ph]];
  M. Cacciari, G.P. Salam and G. Soyez, http://fastjet.fr/

\bibitem{MadGraph}
  F.~Maltoni and T.~Stelzer,
  JHEP {\bf 0302}, 027 (2003);
  [arXiv:hep-ph/0208156];
  J.~Alwall {\it et al.},
  JHEP {\bf 0709}, 028 (2007)
  [arXiv:0706.2334 [hep-ph]].
  
\bibitem{MLM}
  M.~L.~Mangano, M.~Moretti, F.~Piccinini and M.~Treccani,
  JHEP {\bf 0701}, 013 (2007).
  [arXiv:hep-ph/0611129].
  
\bibitem{Chatrchyan:2011jx}
  S.~Chatrchyan {\it et al.}  [CMS Collaboration],
  arXiv:1103.4279 [hep-ex].
  
\bibitem{RPVsub}
  J.~M.~Butterworth, J.~R.~Ellis, A.~R.~Raklev and G.~P.~Salam,
  Phys.\ Rev.\ Lett.\  {\bf 103}, 241803 (2009).
  [arXiv:0906.0728 [hep-ph]].

  
\bibitem{boostedhiggs}
  J.~M.~Butterworth, A.~R.~Davison, M.~Rubin and G.~P.~Salam,
  Phys.\ Rev.\ Lett.\  {\bf 100}, 242001 (2008);
 G.~D.~Kribs, A.~Martin, T.~S.~Roy and M.~Spannowsky,
  Phys.\ Rev.\  D {\bf 81}, 111501 (2010).
  [arXiv:0912.4731 [hep-ph]].


    
\bibitem{Strassler:2008bv}
  M.~J.~Strassler,
  arXiv:0801.0629 [hep-ph].

\end{thebibliography}
\end{document}